\documentclass[preprint]{elsarticle}
\usepackage{graphicx}
\usepackage{epstopdf}
\usepackage[figuresright]{rotating}
\usepackage{amsmath}
\usepackage{amssymb}
\usepackage{boxedminipage}

\newcommand{\beq}{\begin{equation}}

\newcommand{\eeq}{\end{equation}}
\newcommand{\bpm}{\begin{pmatrix}}
\newcommand{\epm}{\end{pmatrix}}
\newcommand{\bea}{\begin{eqnarray}}
\newcommand{\eea}{\end{eqnarray}}

\newcommand{\ignore}[1]{}

\begin{document}
\begin{frontmatter}
\title{Strongdeco: Expansion of analytical, strongly 
correlated quantum states into a many-body basis}

\author[icfo,ecm]{Bruno Juli\'a-D\'{\i}az} 
\author[icfo]{Tobias Gra\ss}

\address[icfo]{ICFO – The Institute of Photonic Sciences, 08860 Castelldefels, Spain}
\address[ecm]{Dept. d'Estructura i Constituents de la Mat\`eria,
  Universitat de Barcelona, 08028, Spain}

\begin{abstract}

We provide a Mathematica code for decomposing strongly correlated
quantum states described by a first-quantized, analytical wave function 
into many-body Fock states. Within them, the single-particle occupations 
refer to the subset of Fock-Darwin functions with no nodes. Such states, 
commonly appearing in two-dimensional systems subjected to gauge fields, 
were first discussed in the context of quantum Hall physics and 
are nowadays very relevant in the field of ultracold quantum gases. 
As important examples, we explicitly apply our decomposition scheme 
to the prominent Laughlin and Pfaffian states. This allows for easily 
calculating the overlap between arbitrary states with these highly 
correlated test states, and thus provides a useful tool to classify
correlated quantum systems. Furthermore, we can directly read off 
the angular momentum distribution of a state from its decomposition. 
Finally we make use of our code to calculate the normalization 
factors for Laughlin's famous quasi-particle/quasi-hole excitations, 
from which we gain insight into the intriguing fractional behavior 
of these excitations.
\end{abstract}
\end{frontmatter}


\noindent{\bf Program Summary}

\noindent{\it Title of program:} Strongdeco\\
{\it Catalogue identifier:}\\
{\it Program summary URL:} http://cpc.cs.qub.ac.uk/summaries\\
{\it Program available from:} CPC Program Library, Queen's University of Belfast, N. Ireland\\
{\it Operating systems:} Linux, Windows, Mac\\ 
{\it Programming language used:} Mathematica\\
{\it Number of bytes in distributed program, including test code and 
documentation:  }\\
{\it Distribution format:} .nb\\
{\it Nature of Problem:} Analysis of strongly correlated quantum states\\
{\it Method of Solution:} The program makes use of the tools developed 
in mathematica to deal with multivariate polynomials to decompose analytical 
strongly correlated states of bosons and fermions into a standard many-body 
basis. Operations with polynomials, determinants and permanents are the 
basic tools. \\

\section{Introduction}

In recent years there has been a growing interest in the study of 
strongly correlated quantum states and their possible realization 
in two-dimensional (2D) systems of ultracold 
atoms~\cite{Lewenstein07,Bloch08}. Such states, which were first 
postulated when studying the dynamics of electrons subjected to
strong magnetic fields, can also be produced in systems of neutral
atoms subjected to so-called artificial gauge fields. One of the 
first examples was obtained by rotating a 2D atomic cloud such that 
the centrifugal force on the atoms mimics the Lorentz force which a 
charged particle would experience in the presence of a constant 
magnetic field perpendicular to the 
system~\cite{Wilkin:2000,fetter,cooper,Dagnino:2009}. The main 
drawback of this approach is that large rotations are needed in order 
to observe strongly correlated states such as the 
Laughlin~\cite{Laughlin:1983}, while it is difficult to stabilize 
in this fast rotating regime. For this reason, the Laughlin state 
has not yet been engineered in the pioneering
experiments ~\cite{Bretin:2004,schweikhard}. Recently, Roncaglia 
{\textit et al.} have proposed an alternative experiment to avoid the 
instability difficulty by using a Mexican-hat trap~\cite{roncaglia}.

Other very promising proposals to overcome this problem come from 
quantum optics and consider the coupling of the atoms to one or 
several laser fields. These make the atoms experience a Berry 
phase~\cite{oehberg,Dalibard:2011,ours11,ours2,levi}, which, due to 
the mathematical equivalence between geometric phases and 
external gauge fields, can then be interpreted as if it were  
due to the presence of an external gauge field. The experimental 
realization of such an artificial gauge field has already been 
achieved~\cite{lin}.

An important motivation to study these systems is the 
possibility of producing strongly correlated quantum states and 
quasi-particle/quasi-hole excitations which are neither 
described by fermionic nor by bosonic commutation laws. The 
latter are expected to have strong impact in the context of 
anyonic quantum computation~\cite{nayak}. One thus needs to 
quantify the strongly correlated states produced by different 
proposals and the properties of their quasi-particle/quasi-hole 
excitations.

A common way to study such systems theoretically is by means 
of exact diagonalization of small-sized systems~\cite{Wilkin:2000,Dagnino:2009}.
The usual methodology is to employ Fock-Darwin (FD) wave functions, 
which describe single particles with fixed angular momentum in 
a fixed Landau level, and which are the eigenfunctions of a 2D 
non-interacting system with a perpendicular magnetic field in 
the symmetric gauge. The many-body basis for bosons (fermions) 
is then built up by symmetric (antisymmetric) combinations of 
the FD states. While this basis is practical for definite 
calculations, many relevant states in the literature have been 
found by proposing a first-quantized, analytic wave function. 
Here the Laughlin~\cite{Laughlin:1983}, the Pfaffian, also 
called Moore-Read~\cite{Moore:1991}, or the Laughlin 
quasi-particle states~\cite{Popp:2004} are the most prominent 
examples. Translating the first-quantized wave functions into 
the language of second quantization, however, turns out to be 
a hard task~\cite{takano,mitra}.

In this paper, we present a computer code which achieves 
this goal for arbitrary states described by an analytic function, 
and thus provides practitioners of this field with a simple 
and yet powerful tool to quantify the degree of correlation 
by examining its expansion into an independent particle motion 
basis. The code is written in Mathematica~\cite{Mathematica}, 
which is a computer language specially suited for symbolic 
evaluation. 

We begin by presenting the first-quantized expression of the 
most important strongly correlated states in Sect.~\ref{sec:st}.
In Sect.~\ref{sec1}, we briefly describe and construct the 
many-body basis into which we then decompose the states in 
Sect.~\ref{sec:deco}. Finally, in Sect.~\ref{sec:app}, we 
consider two applications which can be tackled making use of 
the described decomposition scheme. The most relevant routines 
contained in {\bf Strongdeco.nb} are explained within the 
text, a brief description of all routines is given in the Appendix.

\section{Analytical strongly correlated states}
\label{sec:st}

Strongly correlated states in 2D systems exposed to a gauge field 
are usually studied in the regime where all particles occupy 
the lowest Landau level. The Hilbert space of an $N$-body system 
in this regime can be represented by wave functions of the form
\begin{equation}
\Psi(z_1,\dots,z_N) = {\cal{N}} f(z_1,\dots,z_N) {\mathrm e}^{-\sum \mid
z_i\mid^2/ 2
\lambda_{\perp}^2}\, ,
\label{form}
\end{equation}
where $z_i=(x_i+i y_i)/\lambda_{\perp}$, ${\cal N}$ is a normalization 
constant, and $f$ is a polynomial in its arguments $z_i$. The typical 
length scale of the system is given by $\lambda_{\perp}$. The most 
famous wave function of this form is the Laughlin 
function~\cite{Laughlin:1983}:
\begin{equation}
\Psi_{\cal L}(z_1,\dots,z_N) ={\cal{N}_{\cal L}} f_m(z_i,\dots,z_N)
{\mathrm e}^{-\sum \mid z_i\mid^2/ 2
\lambda_{\perp}^2}\, ,
\label{laughwave}
\end{equation}
with ${\cal N}_{\cal L}$ a normalization constant, and
\begin{equation}
f_m(z_i,\dots,z_N) \equiv \prod_{i<j}(z_i-z_j)^m\,,
\end{equation}
where $m$ is an integer directly related to the filling 
factor $\nu=1/m$ of the lowest Landau level. Originally 
intended to describe electrons, this wave function had 
to be antisymmetric, restricting $m$ to odd numbers. However, 
as shown for instance in Ref.~\cite{Wilkin:2000}, also the 
ground state of a two-dimensional system of rotating bosons 
with contact interaction is, for certain values of the angular 
rotation, described by the Laughlin state, if $m$ is taken as an 
even integer~\cite{Cooper:2001,Regnault:2003}. One important 
property of the Laughlin wave function is that $f_m$ is a 
homogeneous polynomial. Its degree determines the well-defined
total angular momentum of the system, given by $L=\frac{1}{2}mN(N-1)$.

Besides the Laughlin state, other states of the form given 
by Eq.~(\ref{form}) show up as the ground state of a rotating 
Bose gas, if we vary the rotation frequency~\cite{Wilkin:2000,Popp:2004,Roncaglia:2010}.
For a broad range of rotation frequencies, for instance, a large 
overlap is found with the so-called Pfaffian state, which has 
$L=N(N-2)/2$ for even $N$, and $L=(N-1)^2/2$ for odd $N$. It 
explicitly reads, 
\beq
\Psi_{\cal P}= {\cal N}_p  {\rm Pf}([z])  \prod_{i<j}(z_i-z_j)  \ ,
\label{pfaf}
\eeq
with ${\cal N}_p$ a normalization coefficient and, 
\beq
{\rm Pf}([z])= {\cal A} \left[ {1 \over (z_1-z_2)} {1 \over (z_3-z_4)} \cdots {1
    \over (z_{N-1}-z_N)}\right]{\mathrm e}^{-\sum \mid z_i\mid^2/ 2
\lambda_{\perp}^2}\, ,
\eeq
where ${\cal A}$ is an antisymmetrizer of the product. As explained 
in Ref.~\cite{Wilkin:2000}, the Pfaffian state can also be computed 
as, 
\beq
\Psi_{\cal P}= 
{\cal S} 
\prod_{i<j \in \sigma_1} (z_i-z_j)^2
\prod_{k<l \in \sigma_2}(z_k-z_l)^2 \;
{\mathrm e}^{-\sum \mid z_i\mid^2/ 2\lambda_{\perp}^2}\, ,
\label{conje}
\eeq
where $\sigma_1$ and $\sigma_2$ are two subsets containing $N/2$ 
particles each if $N$ is even, and $(N+1)/2$ and $(N-1)/2$ if 
$N$ is odd. ${\cal S}$ symmetrizes the expression. 

Another relevant state is the Laughlin quasi-particle state,
\beq
\Psi_{{\cal L}qp}(\xi,\xi^*)= 
{\cal N}_{qp}(\xi,\xi^*) {\mathrm e}^{-\sum \mid z_i\mid^2/ 2
\lambda_{\perp}^2} \prod_{i\leq N} (\partial_{z_i}-\xi) f_m\,,
\label{lqp}
\eeq
where $\xi$ represents the position of the quasiparticle. If 
we pin the quasi-particle to the origin, the Laughlin 
quasi-particle state has a definite angular momentum 
$L=\frac{1}{2}mN(N-1)-N$, and, as shown in Ref.~\cite{Popp:2004} 
for $m=2$ and $N=4$, its overlap with the ground state of 
rotating ultracold atoms is fairly large in certain regions 
of the rotation.

The analog of Laughlin's quasi-particle state is the quasi-hole 
state, 
\beq
\Psi_{{\cal L}qh}(\xi,\xi^*)=
{\cal N}_{qh}(\xi,\xi^*){\mathrm e}^{-\sum \mid z_i\mid^2/ 2
\lambda_{\perp}^2} \prod_{i\leq N} (z_i-\xi) f_m\,,
\label{lqh}
\eeq
with an increased angular momentum 
$\frac{1}{2}mN(N-1) \leq L\leq \frac{1}{2}mN(N-1)+N$. An interesting 
property of quasiparticles and quasiholes is their anyonic 
nature~\cite{arovas} and fractional charge. Note that in the 
case of an electroneutral system, one may define the analog 
of a charge by looking at the Berry phase a particle acquires 
when moving in the presence of the artificial gauge field.

\section{The many-body basis}
\label{sec1}

After defining in the previous section the general structure 
for all wave functions of interest, we now construct the 
many-body basis into which we wish to decompose these states. 
A convenient choice to span the Hilbert space of the many-body 
system are the eigenfunctions of the non-interacting problem, 
i.e. the FD wave functions,
\beq
\phi_{\ell} (z) = 
{z^\ell  \over  \sqrt{\pi \ell!}} 
{1\over \lambda_\perp^{\ell + 1}}
\;
e^{-|z|^2/(2\lambda_\perp^2)} \,, \quad \ell=0,\dots,\infty \, ,
\label{fd}
\eeq
where we have restricted ourselves to the lowest Landau level.
These states satisfy, 
\beq
\int_{-\infty}^\infty \,dx 
\int_{-\infty}^\infty \,dy\,
\phi^*_{\ell}(z) \phi_{\ell'}(z) =\delta_{\ell,\ell'} \,.
\eeq
The many-body basis can be generated by considering products of 
the FD functions, which in the case of bosons have to be combined 
in a symmetric way, while antisymmetric combinations must be 
constructed in the case of fermions. Here, we will concentrate 
on the bosonic case, but with only slight modifications which 
are explicitly shown in the code file, fermionic systems can be 
treated in the same way. For the bosonic system we write the 
many-body state as, 
\beq
\{\ell_1, \ell_2, \dots, \ell_N\} \equiv 
{\cal S} \;\left[
\phi_{\ell_1}(z_1)
\phi_{\ell_2}(z_2)
\dots
\phi_{\ell_N}(z_N)\right]
\eeq
where ${\cal S}$ symmetrizes over the $N$ particles. These states 
are called permanents, which are the bosonic analog of the Slater
determinants, with the difference that all terms have a positive 
sign. Without loss of generality we may assume that 
$\ell_1 \leq \ell_2 \leq \cdots \leq \ell_N$. The orthonormality 
of the permanents then reads
\beq
\label{onb}
\{\ell_1, \ell_2, \dots, \ell_N\}\cdot\{\ell_1', \ell_2', \dots, \ell_N'\} =
\delta_{\ell_1,\ell_1'}\delta_{\ell_2,\ell_2'} \cdots \delta_{\ell_N,\ell_N'}\,.
\eeq

For simplicity we will from now on set the scale factor $\lambda_\perp=1$, 
and suppress the exponential term which is common to all $N$-body 
states, and, as an overall Gaussian, fixes the center of mass to the 
origin. We can then simplify the problem to dealing with permanents 
of the form, 
\beq
{\cal S}\left[z_1^{\ell_1} z_2^{\ell_2}\dots  z_N^{\ell_N} \right]\,.
\eeq
From Eq.~(\ref{onb}) follows that, for a given $N$, all states of a 
fixed total angular momentum $L=\sum_{i=1}^N \ell_i$ form a subspace 
which is orthogonal to the subspace with total angular momentum 
$L' \neq L$. We can therefore perform the decomposition independently 
in each subspace, and thus restrict ourselves to a subspace with 
fixed $L$. Its basis (up to normalization factors and the overall 
exponential term) can be constructed through the command, 

\vspace{10pt}
\begin{boxedminipage}{0.8\textwidth}
\begin{verbatim}
ConjS[na_, L_] := Module[{poty, dimy},
  poty = Pots[na, L];
  dimy = Dimensions[poty][[1]];
  Table[Perm[na, poty[[i]]], {i, 1, dimy}]]
\end{verbatim}
\end{boxedminipage}
\vspace{10pt}

\noindent which makes use of the function {\bf Perm}~\cite{Eric}, 
that builds the appropriate permanent, and of {\bf Pots[N,L]}, which 
constructs the set of indexes ${\ell_1, \dots \ell_N}$ for a given 
$N$ and $L$, represented by {\bf na} and {\bf L} in the code,

\vspace{10pt}
\begin{boxedminipage}{0.94\textwidth}
\begin{verbatim}
cc[0] = 0;
tab[n_, l_] := 
Table[{cc[i], cc[i - 1], 
If[i == 1, l, (l - Sum[cc[j], {j, 0, i - 1}])/2]}, 
{i, 1, n - 1}];
Pots[na_, L_] := If[na == 2, Table[{i, L - i}, {i, 0, L/2}],
Module[{pat},
Clear[pat];
pat[na] = Join[Table[
cc[i], {i, 1, na - 1}], {L - Sum[cc[i], {i, 1, na - 1}]}];
pat[a_] := Table[pat[a + 1], Evaluate[tab[na, L][[a]]]];
Flatten[pat[1], na - 2]]]
\end{verbatim}
\end{boxedminipage}

\vspace{10pt}
\noindent For instance, for $N=4$ and $L=2$ we have
~\footnote{Note that the state $\{1,1,0,0\}$ is equivalent to $\{0,0,1,1\}$
  due to the symmetrization of the states.}, 
\begin{verbatim}
Pots[4,2]={{0,0,0,2},{0,0,1,1}}
\end{verbatim}
and correspondingly, 
\begin{verbatim}
ConjS[4, 2]=
{6 z[1]^2+6 z[2]^2+6 z[3]^2+6 z[4]^2,
4 z[1] z[2]+4 z[1] z[3]+4 z[2] z[3]
+4 z[1] z[4]+4 z[2] z[4]+4 z[3] z[4]}
\end{verbatim}
As can be seen in this example, due to multiple occupation 
of the same single-particle state, some of the permutations 
contributing to the symmetrized wavefunction are described 
by the same monomials which thus have prefactors given 
by the factorial of the number of permutations. These 
factors need to be taken into account to correctly normalize 
the many-body states, and can be obtained through, {\bf nami[N, L]}, 
which gives a table with the same ordering as {\bf Pots} or 
{\bf ConjS}, for our previous example, {\bf nami[4,2]=\{6,4\}}, 
as could be inferred from the obtained expressions. 

\vspace{10pt}
\begin{boxedminipage}{0.94\textwidth}
\begin{verbatim}
nami[na_, L_] := Module[{potty, pp, inde, ta},
  potty = Pots[na, L];
  pp = Dimensions[potty][[1]];
  inde = Table[Complement[potty[[i]]], {i, 1, pp}];
  ta = Table[Table[Count[potty[[i]], inde[[i, j]]], 
{j, 1, Dimensions[inde[[i]]][[1]]}], {i, 1, pp}];
  Table[ Product[ta[[i, j]]!, 
{j, 1, Dimensions[ta[[i]]][[1]]}]   , {i, 1, pp}]]
\end{verbatim}
\end{boxedminipage}
\vspace{10pt}

\noindent Once these factors are known it is easy to build 
the normalization coefficient by looking into the prefactors 
in the Fock-Darwin states Eq.~(\ref{fd}), $1/\sqrt{\pi \ell!}$. 
The function {\bf tip[N, L]} gives the normalization 
coefficients. Their explicit coding is,

\vspace{10pt}
\begin{boxedminipage}{0.95\textwidth}
\begin{verbatim}
tip[na_, L_] := Module[{potty, nimy},
potty = Pots[na, L];
nimy = nami[na, L];
Table[Sqrt[nimy[[i]]]Sqrt[Product[Pi Gamma[potty[[i, jj]]+1], 
{jj, 1, na}]  ], {i, 1, Dimensions[nimy][[1]]}]]
\end{verbatim}
\end{boxedminipage}

\section{Decomposition of the states}
\label{sec:deco}

All states described in Sect.~\ref{sec:st}, are, up to 
the common exponential factor, polynomials in the $z$ 
variables. To write down the states in terms of the 
many-body ones, we can suppress the exponential and work 
out the decomposition of the polynomial in terms of the 
permanents. While the Laughlin and the Pfaffian state have 
a definite total angular momentum, for the quasi-hole and 
quasi-particle states this is only true if we fix the 
position of the quasi-particle to the origin $\xi=0$. Otherwise, 
we must first sort the polynomial by the different contributions
with a definite order in $z$, and can then proceed, for each 
contribution separately, in the way described here, where 
we assume an analytical state, $\Psi(z_1,z_2,\dots,z_N)$ 
with fixed $N$ and $L$. We are looking for an expansion of 
the form, 
\beq
\label{dec}
\Psi(z_1,z_2,\dots,z_N) = \sum_{j=1}^{n_D}  C_j   \{
\ell_{1,j},\ell_{2,j},\dots, \ell_{N,j}\}
\eeq
where $n_D$ is the total size of the many-body basis, 
which can be computed as $n_D={\rm Dimensions[PotsN}[N,L]][[1]]$. 
To get a feeling of how this grows with $N$ and $L$ the dimension 
of these spaces for the $L$ corresponding to the Laughlin wave 
functions are, $n_D=7, 34, 192, 1206, 8033, 55974$ for
$N=3,4,5,6,7$ and 8, respectively.

To decompose a polynomial into these states, we have to find 
the monomials which correspond to a given Fock state and read 
out their coefficients. Since we know that the polynomial is 
symmetric (antisymmetric) under exchange of two coordinates, 
it is sufficient to find only one monomial contributing to a given
Fock state, as all the others must have the same coefficient (up to a sign in
the antisymmetric case). This can be achieved by taking derivatives:
\beq
\partial_{z_1}^{\ell_{1,j}} \cdots \partial_{z_N}^{\ell_{N,j}}
\Psi(z_1,z_2,\dots,z_N) |_{z_1=0,\cdots,z_N=0}
= c_j,
\label{eq15}
\eeq
where $c_j$ is not yet the coefficient $C_j$ in Eq. (\ref{dec}), but is directly
related to it through the normalization procedure described in the previous
section. Hereby, we have to take into account that an additional factor
$\prod_{i=1}^{N} \ell_{i,j}!$  occurs through the derivatives. Thus we
obtain
\beq
C_j = c_j \left(P \prod_{i=1}^{N} \ell_{i,j}! 
\ \pi \right)^{-1/2} \equiv
\alpha_j c_j,
\eeq
where the $P$ is the factorial of permutations leading to the same expression,
obtained by {\bf nami[N,L]}. We thus see that $\alpha_j$ equals the inverse
of the $j$th component of ${\bf tip[N,L]}$. The decomposition of, for instance,
the Laughlin wavefunction can therefore be obtained by the following piece of
code:

\vspace{10pt}
\begin{boxedminipage}{0.97\textwidth}
\begin{verbatim}
DDecoLaug[na_,nu_] := 
 Module[{Lmin, Lmax, state, base, dim, factors, d, prf, outp},
  Lmin = na (na - 1);
  Lmax = na (na - 1);
  state = Laughlin[na,nu];
  base = Flatten[Table[Pots[na, i], {i, Lmin, Lmax}], 1];
  dim = Dimensions[base][[1]];
  factors = Flatten[Table[tip[na, i], {i, Lmin, Lmax}], 1];
  d[0] = state;
  prf = Table[
    For[i = 1, i < na + 1, i++,
     d[na] = 0;
     d[i] = D[d[i - 1], {z[i], base[[j, i]]}];
     d[i] = d[i] /. z[i] -> 0;
     If[d[i] == 0, Break[]]
     ];
    d[na]/factors[[j]],
    {j, 1, dim}];
  outp = prf/Sqrt[prf.prf]]
\end{verbatim}
\end{boxedminipage}
\vspace{10pt}

\noindent Here, {\bf Laughlin[N,nu]} describes the Laughlin wavefunction 
for $N$ particles at filling $\nu$. For even $1/\nu$, this is a symmetric 
function describing bosons, while odd values yield an antisymmetric 
function for fermionic systems. In principle, we can use the code 
for both the symmetric and the antisymmetric case. In the latter, 
however, it is convenient to exclude states with multiple occupied 
single-particle levels from the basis, as they obviously make no 
contribution. This can be done by replacing {\bf Pots[N,L]} by its 
fermionic analogue {\bf PotsF[N,L]} defined in the code file. 
Consequently, we will also have to replace {\bf tip[N,L]} by 
{\bf tipF[N,L]}.

An alternative way to achieve the decomposition is by
means of a particular, built-in Mathematica function, 
{\bf PolynomialReduce}. This function provides the decomposition 
of a given multivariate polynomial in terms of a set of 
polynomials. The code for decomposing the bosonic Laughlin 
state then reads

\vspace{10pt}
\begin{boxedminipage}{0.97\textwidth}
\begin{verbatim}
LaugDeco[na_,nu_]:=Module[{state, base, symb, laur, prf, outp},
  state = Laughlin[na,nu];
  base = ConjS[na, na (na - 1)];
  symb = Table[z[i], {i, 1, na}];
  laur = PolynomialReduce[state, base, symb];
  If[laur[[2]] != 0, Print["Problem in reduction"]];
  prf = laur[[1]] tip[na, na (na - 1)];
  outp = prf/Sqrt[prf.prf];
  outp]
\end{verbatim}
\end{boxedminipage}
\vspace{10pt}

\noindent For most states that we have considered, the 
decomposition by means of derivatives is faster. However, 
making use of {\bf PolynomialReduce} turns out to be 
quicker for the fermionic Laughlin state as well as 
for quasiparticle excitations.

\begin{figure}[h]
\fbox{\parbox{1.0\textwidth}{\includegraphics[width=12.cm]{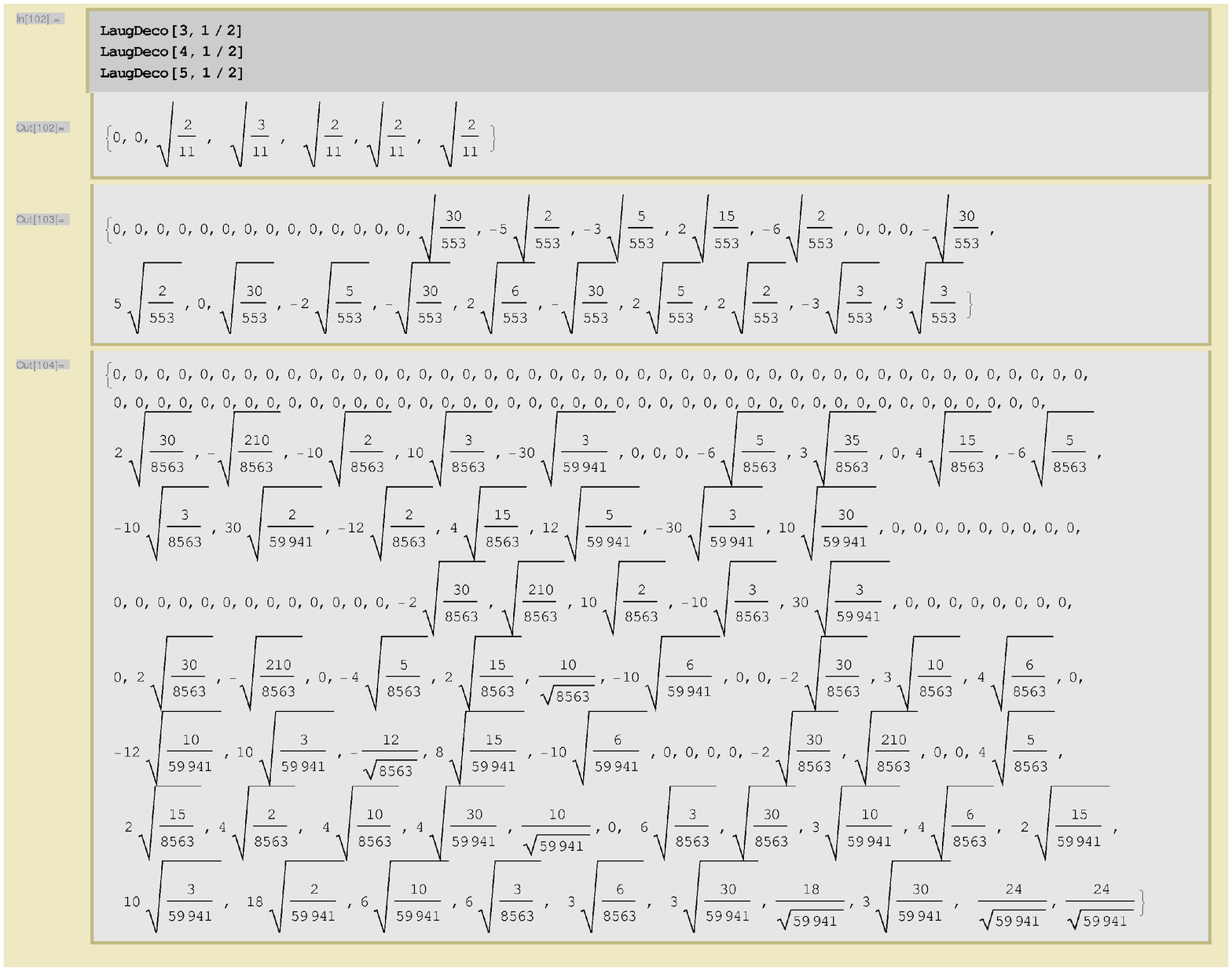}}}
\caption{Snapshot of the code where the decomposition of the 
Laughlin state with $\nu=1/2$ is obtained for $N=3$, $N=4$ and $N=5$. }
\protect\label{figy}
\end{figure}

In figure~\ref{figy}, a snapshot of the code 
for the decomposition of the Laughlin state is provided 
for $N=3$, $N=4$ and $N=5$. The code has been tested for $N\leq 7$ 
on a laptop running on linux with 1Gb of RAM memory. A listing 
of the different commands defined in {\bf Strongdeco.nb} is 
provided in the Appendix. The notebook is provided with some 
examples built-in inside. 

\section{Applications}               
\label{sec:app}

Finally, we make use of the presented decomposition scheme, 
and consider as examples three problems which might be 
tackled with the given code.

\subsection{Wave-function overlaps}

The decomposition achieved by our code turns out to be very
useful if the state, e.g. the eigenstates of a certain problem 
or the evolved state at a given time, of a system is known 
in the many-body basis. This is the case if a system is 
studied via exact diagonalization ~\cite{Wilkin:2000,Dagnino:2009,ours11,ours2}. 
It is then customary to ask whether, and to which degree, the 
obtained state resembles one of the well-known strongly
correlated states described in Sect.~\ref{sec:st}. To answer 
this question one has to calculate the overlap between both 
states, which can either be done by straightforwardly expressing 
the many-body state in first quantization and then evaluating the 
overlap integrals. For reasonable system sizes of $N \geq 3$, 
the latter is a very lengthy task. The second possibility
consists of finding an expression of the analytic wave function 
in terms of the many-body basis, which then reduces the overlap 
calculation to a simple scalar multiplication of two vectors. 
In this case, the first step is the non-trivial one, but it is 
directly achieved by the code we have presented here.

\subsection{Angular-momentum distribution}

As a second application, one can consider a system which is known 
to have eigenstates of the form given by Eq.~(\ref{form}). Many of 
the properties of such states are better computed by first transforming 
it into the independent motion basis. A clear example is the calculation 
of the angular momentum distribution of the state, from which one also 
gains insight into its one-body density matrix and other correlation 
functions. For the fermionic Laughlin state this problem has been 
considered in Ref.~\cite{mitra}, where exact results are obtained by 
calculating the density and then extracting the angular momentum 
distribution. This method, however, fails for systems larger than 
$N=3$, for which MonteCarlo methods have been applied. By means of 
our code, we are able to reproduce these results by decomposing the 
Laughlin state into a basis from which the angular momentum of each 
particle can be directly read off. It is straightforward to go beyond the
analytical results of Ref. \cite{mitra}. As an example, 
we give in Table~\ref{ad} the angular momentum distributions for 
the fermionic Laughlin state at $m=3$ and the bosonic Laughlin
state at $m=2$ of a system with $N=4$ particles.

\begin{table}
\begin{center}
\begin{tabular}{|l|cccccccccc|}
 \hline
$\ell=$ & $0$ & $1$ & $2$ & $3$ & $4$ & $5$ &
$6$ & $7$ & $8$ & $9$ \\ 
\hline
& \multicolumn{10}{l|}{}\\
$m=2$ & $\frac{257}{553}$ & $\frac{264}{553}$& $\frac{303}{553}$&
$\frac{446}{553}$& $\frac{447}{553}$& $\frac{330}{553}$& $\frac{165}{553}$&
0& 0& 0\\
& \multicolumn{10}{l|}{}\\
$m=3$   & $\frac{185}{706}$ & $\frac{185}{706}$& $\frac{209}{706}$&
$\frac{321}{706}$& $\frac{417}{706}$& $\frac{465}{706}$& $\frac{455}{706}$&
$\frac{339}{706}$& $\frac{186}{706}$& $\frac{62}{706}$\\
& \multicolumn{10}{l|}{}\\
\hline
\end{tabular}
\caption{\label{ad} Angular-momentum distribution for Laughlin states of $N=4$
particles for bosons ($m=2$) and fermions ($m=3$).}
\end{center}
\end{table}

\subsection{Fractional charge of excitations}

Another useful application is the calculation of the 
normalization factor for a state of the form (\ref{form}). 
As explained in Ref.~\cite{wen}, the normalization factors 
${\cal N}_{qh}(\xi,\xi^*)$ and ${\cal N}_{qp}(\xi,\xi^*)$ of the 
quasi-particle in Eq.~(\ref{lqp}) and the quasi-hole state in
Eq.~(\ref{lqh}), contain information about the Berry phase 
${\bf a}\cdot\mathrm{d}{\bf x}$ which these excitations acquire 
during an adiabatic movement:
\begin{equation}
 a_{\xi} \equiv \frac{i}{2} \partial_{\xi} \mathrm{ln} {\cal N}(\xi,\xi^*),
\end{equation}
with $\partial_{\xi}\equiv\frac{1}{2}(\partial_x -i \partial_y)$ 
and $a_{\xi}\equiv\frac{1}{2}(a_x-ia_y)$. With this expression one 
may consider the phase picked up by the quasi-particle when it 
is moved adiabatically around a closed loop, which is given 
by $I_a \equiv \oint\mathrm{d}{\bf x}\cdot {\bf a}$. Comparing 
this with the phase a moving particle acquires in the system, 
$I_A \equiv \oint \mathrm{d}{\bf  x}\cdot{\bf A}$, 
where ${\bf A}$ is the external gauge potential, one is able to deduce 
the fractional charge and fractional statistics of the quasi-particles. 
If both loop integrals are equal, i.e. $\eta\equiv I_a/I_A=1$, the 
quasi-particle behaves like a normal particle, while the mismatch 
by a fractional factor, i.e. $\eta=1/p$ with $p$ an integer, allows to 
interpret the quasi-particle as a ``fractional''particle.

The difficulty of this analysis lies in the calculation of 
the normalization factors. For the Laughlin state one usually 
circumvents this by applying the Plasma analogy~\cite{Laughlin:1983,arovas} 
to determine the normalization factor of the corresponding 
quasi-holes and quasi-particles, avoiding the explicit calculation. 
It is found that $p=m$, i.e. the fractional behavior of the 
excitation follows from the fractional filling of the lowest 
Landau level. The plasma analogy, however, is not applicable 
to all the relevant states, which exhibit such anyonic excitations, 
but are different from the Laughlin. Calculating the normalization 
factors by direct integration is much too complicated, even for 
systems of only a few particles. However, by transforming the 
corresponding quasi-hole or quasi-particle state into the 
many-body basis by means of our code, the normalization factors are 
obtained by simply taking the scalar product of the decomposition vector. 

As an example, we calculate $\eta_{qh}$ and $\eta_{qp}$ for
quasi-holes and quasi-particles in small Laughlin systems. 
First, we notice that the polynomial in Eqs.~(\ref{lqp}) and 
(\ref{lqh}) contains terms of different order in $z$. It is 
therefore necessary to separate the $N$ contributions with 
fixed angular momentum, and then apply the decomposition explained 
in Sect. \ref{sec:deco} to each of them. Then we can write the 
normalization factor as a polynomial in $\xi$ and $\xi^*$. For 
instance, a the wavefunction of a quasi-hole in the $m=2$ Laughlin 
state of $N=5$ bosons, is found to have the following normalization factor:
\begin{eqnarray}
 C \propto 1 + 0.477 |\xi|^2 + 0.117 |\xi|^4 + 0.0211 |\xi|^6 + 0.00334 |\xi|^8
+ 0.000668 |\xi|^{10},
\end{eqnarray}
which, even in high orders of $\xi$, agrees reasonably well with 
the prediction by the plasma analogy, according to which 
$C \propto \exp(\frac{1}{m}|\xi|^2)$.

However, from Table~\ref{fc} showing the results for different 
numbers of particles, we find that, especially for the quasi-particle 
excitation, significant deviations occur for very small systems 
$(N=4)$. For $N=6$, the mismatch is almost cured, and the agreement 
with the plasma analogy prediction is better than 3\% for both 
quasi-holes and quasi-particles.

\begin{table}
\begin{center}
\begin{tabular}{|l|c|c|}
 \hline
$N$ & $\eta_{qh}$ & $\eta_{qp}$ \\
 \hline
4 & 0.473 & 0.354 \\ 
5 & 0.477 & 0.438 \\
6 & 0.487 & 0.501 \\
\hline
\end{tabular}
\caption{\label{fc} Fractional charge of quasi-holes and quasi-particles 
in a bosonic Laughlin system at filling $\nu=1/2$ for different number 
of particles $N$.}
\end{center}
\end{table}

\section{Conclusion}

We have presented a code which, by means of the symbolic 
package Mathematica, analytically  decomposes relevant 
analytical, strongly-correlated many-body states into 
the many-body basis built up by single-particle angular 
momentum eigenstates. This basis is commonly used to describe 
2D quantum systems subjected to gauge fields, and thus the 
described decomposition is a useful tool for calculating 
the overlap of different states with famous test states 
like the Laughlin state, the Pfaffian state and many others. 
It also allows for studying the angular-momentum distribution 
of strongly correlated states, which can be related to the 
ground-state density of the system, or normalization constants 
of the states. The latter has been shown to provide insights 
into the fractional character of quasiparticle excitations.

\section*{Acknowledgments}
The authors thank Frank Tabakin and Nuria Barber\'an for a 
careful reading of the manuscript and useful suggestions. 
B.~J.-D. is supported by a Grup Consolidat SGR 21-2009-2013. 
This project is partially supported by Grants No. FIS2008-01661 
(Spain), and No. 2009SGR1289 from Generalitat de Catalunya. 
We acknowledge financial support from the Spanish MEC project 
TOQATA and the ERC Advanced Grant QUAGATUA.

\appendix
\section{List of routines in Strongdeco.nb}

We provide a brief description of the routines, 

\begin{table}
\begin{tabular}{|l|l|}
\hline
Command Name           & Brief Description \\
\hline
\hline
\hline
\multicolumn{2}{|c|}{Many-body states}\\
\hline
{\bf Pots[N, L]}         & Many-body states for $N$ bosons of fixed total $L$\\
{\bf ConjS[N,L]}         & Permanents for $N$ bosons of fixed total $L$\\
{\bf PotsF[N,L]}         & Many-body states for $N$ fermions of fixed total $L$\\
{\bf ConjSF[N,L]}        & Slaters for $N$ fermions of fixed total $L$\\
\hline
\multicolumn{2}{|c|}{Sample states}\\
\hline
{\bf Laughlin[N,$\nu$]}         & Laughlin state of $N$ atoms and filling $\nu$.\\
{\bf Pfaffian[N,$\nu$]}         & Pfaffian state of $N$ atoms using a Laughlin
of filling $\nu$\\
{\bf Conje[N]}         & Pfaffian state for $N$ atoms, filling $1/2$\\
{\bf deltaL2[N,I]}     & Generalized Laughlin state \cite{ours11}  \\
{\bf deltaL4[N,I]}         &Generalized Laughlin state \cite{ours11} \\
{\bf qh[N,$\nu$,$\xi$]}         & Laughlin quasi-hole state at position $\xi$\\
{\bf quah[N,$\nu$]}         & Laughlin quasi-hole state at the center\\
{\bf qp[N,$\nu$,$\xi$]}        & Laughlin quasi-particle state at position
$\xi$ \\
{\bf quap[N]}        & Laughlin quasi-particle state at the center ($\nu=$1/2)\\
{\bf deltaQP2[N,i]}         & Generazlied Laughlin quasi-particle\\
{\bf edge[N]}         & Laughlin ($\nu=1/2$) edge excitation \\
{\bf deltaP2[N,i]}         &Generalized Pfaffian\\
{\bf deltaP4[N,i]}        &Generalized Pfaffian\\
\hline
\multicolumn{2}{|c|}{Normalizations}\\
\hline
{\bf nami[N,L]}            & Repeated terms, bosons\\
{\bf tip[N,L]}            & Normalization coefficient, bosons\\
{\bf tipF[N,L]}           & Normalization coefficient, fermions\\
\hline
\multicolumn{2}{|c|}{Decomposition of states}\\
\hline
\multicolumn{2}{|l|}{Bosons}\\
\hline
{\bf LaugDeco[N,$\nu$]}          &Laughlin decomposition\\
{\bf DDecoLaug[N,$\nu$]}         & Laughlin decomposition using
Eq.~(\ref{eq15})\\
{\bf EdgeDeco[N]}          & Laughlin edge state\\
{\bf Laug2Deco[N]}        & Generalized Laughlin\\
{\bf Laug4Deco[N]}         & Generalized Laughlin\\
{\bf QuasiHDeco[N]}         & Quasihole\\
{\bf DDecoQuah[N]}         & Quasihole using Eq.~(\ref{eq15})\\
{\bf NormQh[N]}         & Quasihole norm\\
{\bf QuasiPDeco[N]}          & Laughlin quasiparticle\\
{\bf DDecoQuap[N]}          &Laughlin quasiparticle using Eq.~(\ref{eq15})\\
{\bf QuasiP2Deco[N]}         & Generalized Laughlin quasiparticle\\
{\bf PfaffDeco[N]}          & Pfaffian\\
{\bf ConjeDeco[N]}         & Pfaffian using Eq.~(\ref{conje})\\
{\bf DDecoConje[N]}          &Pfaffian using Eq.~(\ref{conje}) and
Eq.~(\ref{eq15})\\
{\bf Pfaff2Deco[N]}          & Generalized Pfaffian\\
\hline
\multicolumn{2}{|l|}{Fermions}\\
\hline
{\bf LaugDecoF[N,$\nu$]}         & Laughlin\\
{\bf DDecoLaugF[N, $\nu$]}   & Laughlin using Eq.~(\ref{eq15})\\
{\bf QuasiHDecoF[N,$\nu$]}   &Laughlin quasi hole\\
\hline
\end{tabular}
\caption{List of commands provided in {\bf Strongdeco.nb}.}
\end{table}

\end{document}